\documentclass[preprint,aps,12pt,showpacs,nofootinbib,tightenlines]{revtex4}
\usepackage{mathrsfs}
\usepackage{amsmath}
\usepackage{amssymb}
\usepackage{CJK}
\usepackage{epsfig}
\usepackage{setspace}
\usepackage{graphicx}
\usepackage{subfigure}
\usepackage{slashed}
\let\jnfont=\rm
\def\NPB#1,{{\jnfont { \it Nucl.\ Phys.\ B }}{\bf #1}}
\def\PLB#1,{{\jnfont { \it Phys.\ Lett.\ B }}{\bf #1}}
\def\EPJC#1,{{\jnfont {\it Eur.\ Phys.\ Jour.\ C }}{\bf #1}}
\def\EPL#1,{{\jnfont { \it Europhys.\ Lett\ }}{\bf #1}}
\def\PRD#1,{{\jnfont {\it Phys.\ Rev.\ D }}{\bf #1}}
\def\PRL#1,{{\jnfont { \it Phys.\ Rev.\ Lett\ }}{\bf #1}}
\def\MPLA#1,{{\jnfont{ \it  Mod.\ Phys.\ Lett\ A, }}{\bf #1}}
\def\IJMPA#1,{{\jnfont{ \it  Int.\ J.\ Mod.\ Phys.\ A }}{\bf #1}}
\def\JPG#1,{{\jnfont {\it J.\ Phys.\ G\ }}{\bf #1},}
\def\CTP#1,{{\jnfont{ \it Commun.\ Theor.\ Phys\ }}{\bf #1}}
\def\SCG#1,{{\jnfont{ \it Sci.\  China.\ G\ }}{\bf #1}}
\def\CPL#1,{{\jnfont{ \it Chin.\ Phys.\ Lett\ }}{\bf #1}}
\def\JHEP#1,{{\jnfont {\it JHEP \ }}{\bf #1}}
\def\NPPS#1,{{\jnfont {\it Nucl.\ Phys.\ Proc.\ Suppl\ }}{\bf #1}}

\def\pslash{\rlap{\hspace{0.02cm}/}{p}}
\begin{document}
\def\pslash{\rlap{\hspace{0.02cm}/}{p}}
\def\eslash{\rlap{\hspace{0.02cm}/}{e}}
\title {Higgs boson production and decay at $e^{+}e^{-}$ colliders as a probe of the Left-Right twin Higgs model}
\author{Jinzhong Han$^{1}$}\email{hanjinzhong@zknu.edu.cn}
\author{Shaofeng Li$^{1}$}
\author{Bingfang Yang$^{2}$}\email{yangbingfang@htu.edu.cn}
\author{Ning Liu$^{2}$}\email{wlln@mail.ustc.edu.cn}
\affiliation{\footnotesize $^1$School of Physics and
Electromechanical Engineering, Zhoukou Normal University, Henan, 466001, China\\
$^2$College of Physics and Electronic Engineering, Henan Normal
University, Xinxiang 453007, China
   \vspace*{1.5cm}  }
\begin{abstract}

In the framework of the Left-Right twin Higgs (LRTH) model, we
consider the constrains from the latest search for high-mass
dilepton resonances at the LHC and find that the heavy neutral boson
$Z_H$ is excluded with mass below 2.76 TeV. Under these constrains,
we study the Higgs-Gauge coupling production processes
$e^{+}e^{-}\rightarrow ZH$, $e^{+}e^{-}\rightarrow
\nu_{e}\bar{\nu_{e}}H$ and $e^{+}e^{-}\rightarrow e^{+}e^{-}H$, top
quark Yukawa coupling production process $e^{+}e^{-}\rightarrow
t\bar{t}H$, Higgs self-couplings production processes
$e^{+}e^{-}\rightarrow ZHH$ and $e^{+}e^{-}\rightarrow
\nu_{e}\bar{\nu_{e}}HH$ at $e^{+}e^{-}$ colliders. Besides, we study
the major decay modes of the Higgs boson, namely $h\rightarrow
f\bar{f}$($f=b,c,\tau$), $VV^{*}(V=W, Z)$, $gg$, $\gamma\gamma$. We
find that the LRTH effects are sizable so that the Higgs boson
processes at $e^{+}e^{-}$ collider can be a sensitive probe for the
LRTH model.

\end{abstract}
 \maketitle
\section{ Introduction}
\noindent

The hunt for Higgs bosons is one of the most important goals at the
Large Hadron Collider (LHC). On the 4th of July 2012, CERN announced
that both the ATLAS \cite{ATLAS} and CMS \cite{CMS} experiments
presented very strong evidence for a new Higgs-like boson with mass
around 125 GeV. With the growingly accumulated date, the properties
of this particle are consistent with those of  Higgs boson predicted
by the Standard Model (SM)
 \cite{spin-atlas,spin-cms}. Though the
LHC offers obvious advantages in proving very high energy and very
large rates in typical reactions, the measuring precision will be
restricted due to the complicated background. However, the most
precise measurements will be performed in the clean environment of
the future $e^{+}e^{-}$ colliders, like the International Linear
Collider (ILC) \cite{ILC-report}.

It is well known, the main production processes of the Higgs boson
in $e^{+}e^{-}$ collisions are the Higgs-strahlung process
$e^{+}e^{-}\rightarrow ZH$ and the $WW$ fusion process
$e^{+}e^{-}\rightarrow \nu_{e}\bar{\nu_{e}}H$. The cross section for
the Higgs-strahlung process is dominant at the low energy. For
$\sqrt{s}\geq 500$ GeV, the cross section for the $WW$ fusion  is
dominant. The cross section for the $ZZ$ fusion process
$e^{+}e^{-}\rightarrow e^{+}e^{-}H$ increases significantly with the
center-of-mass (c.m.) energy increasing, and can exceeds that of
$ZH$ production around 1 TeV. These three processes can be well used
to test the Higgs-Gauge couplings.

The large top quark Yukawa coupling is speculated to be sensitive to
new physics, it can be studied through the associated production of
Higgs boson with top quark pairs $e^{+}e^{-}\rightarrow t\bar{t}H$
at the ILC. This study will play an important role for precision
measurements of the top quark Yukawa coupling. In addition, the
Higgs self-coupling is the key ingredient of the Higgs potential and
their measurement is indispensable for understanding the electroweak
symmetry breaking. The Higgs self-coupling can be studied through
the double Higgs boson production processes $e^{+}e^{-}\rightarrow
ZHH$ and $e^{+}e^{-}\rightarrow \nu_{e}\bar{\nu_{e}}HH$ at the ILC.
And many relevant works mentioned above have been extensively
studies in the context of the SM \cite{SM-zh-vvh-eeh} and some new
physics models \cite{np-zh,np-vvh,np-tth}.

As an extension of the SM, the Left-Right twin Higgs (LRTH) model
has been proposed as an alternative solution to the little hierarchy
problem \cite{LRTH-1,LRTH-2}. The idea of twin Higgs similar to that
of little Higgs, in that the SM-like Higgs emerges as a
pseudo-Goldstone boson \cite{little Higgs}. The twin Higgs mechanism
can be implemented in LRTH model with the discrete symmetry being
identified with left-right symmetry. The phenomenology of the LRTH
model has been studied in Refs.
\cite{phenomenology-1,phenomenology-2,phenomenology-3,phenomenology-4,phenomenology-5}.
In the LRTH model, some new particles are predicted and some SM
couplings are modified so that the Higgs properties may deviate from
the SM Higgs boson. So the Higgs boson  processes are ideal ways to
probe the LRTH model at the $e^{+}e^{-}$ colliders. In this paper,
we mainly study the Higgs boson production processes
$e^{+}e^{-}\rightarrow ZH$, $e^{+}e^{-}\rightarrow
\nu_{e}\bar{\nu_{e}}H$, $e^{+}e^{-}\rightarrow e^{+}e^{-}H$,
$e^{+}e^{-}\rightarrow t\bar{t}H$, $e^{+}e^{-}\rightarrow ZHH$ and
$e^{+}e^{-}\rightarrow \nu_{e}\bar{\nu_{e}}HH$. Besides, we consider
the major decay modes of the Higgs boson $h\rightarrow
f\bar{f}$($f=b,c,\tau$), $VV^{*}(V=W, Z)$, $gg$, $\gamma\gamma$ in
the LRTH model.

The paper is organized as follows. In Sec.II we briefly review the
basic content of the LRTH model related to our work. In Sec.III and
Sec.IV we respectively investigate the Higgs boson production and
decay processes, and give the numerical results and discussions.
Finally, we give a short conclusion in Sec.V.

\section{ A brief review of the LRTH model}
Here we will briefly review the ingredients which are relevant to
our calculations, and a detailed description of the LRTH model can
be found in Ref \cite{phenomenology-1}. The LRTH model introduces
the heavy gauge bosons $W^{\pm}_H$ and $Z_H$, the extra Higgs bosons
$\phi^{0}$ and $\phi^{\pm}$, and the top quark partner $T$. The
masses of theses particles are given by:
\begin{eqnarray}
&&M_{W_{H}}^{2}= \frac{1}{2}g^{2}(\hat{f}^{2}+f^{2}\cos^{2}x),\\
&&M_{Z_{H}}^{2}=
\frac{g^{2}+g'^{2}}{g^{2}}(M_{W}^{2}+M_{W_{H}}^{2})-M_{Z}^{2},\\
&&m^{2}_{\phi^0}=\frac{\mu_{r}^{2}f\hat{f}}{\hat{f}^{2}+f^{2}\cos^{2}x}
\left \{\frac{\hat{f}^{2}\left[ \cos x+\frac{\sin x}{x}(3+x^{2})
\right]}{f^{2}\left (\cos x+\frac{\sin x}{x} \right )^{2}}+2\cos
x+\frac{f^{2}\cos^{2}x(1+\cos x)}{2\hat{f}^{2}} \right\},\\
&&M_{T}^{2}=\frac{1}{2}(M^{2}+y^{2}f^{2}+N_{t}),
 \end{eqnarray}
where $g=e/{S_W}$, $g'=e/{\sqrt{\cos2\theta_W}}$,
$S_W=\sin\theta_W$, $\theta_{W}$ is the Weinberg angle,
$x=v/(\sqrt{2}f)$ with $v$ is the electroweak scale, $N_{t} =
\sqrt{(y^{2}f^{2}+M^{2})^{2}-y^{4}f^{4}\sin^{2}2x}$ and the mass
parameter $M$ is the mixing between the SM top quark and its partner
$T$. The Higgs vacuum expectation values (VEVs) $f$, $\hat{f}$ will
be bounded by electroweak precision measurements. If we set $v$ =246
GeV, $f$ and $\hat{f}$ will be interconnected. In addition, the top
Yukawa coupling will also be of order one if $M_{T} \leq f$ and the
parameter $y$ is of order one. The expression form of the couplings
related to our calculations are given as follows
\cite{phenomenology-1,calchep}:
\begin{eqnarray}
&&g^{t\bar{T}Z}_L=\frac{eC_{L}S_{L}}{2C_{W}S_{W}},~~~~~~~~~~~~~~{{g^{t\bar{T}Z}_R=\frac{ef^2x^2S_WC_{R}S_{R}}{2\hat {f}^2C^3_{W}}}}; \\
&&g^{t\bar{T}Z_H}_L=\frac{eC_{L}S_{L}S_W}{2C_{W}\sqrt{\cos2\theta_W}},~~~~~g^{t\bar{T}Z_H}_R=-\frac{eC_{R}S_{R}C_W}{2S_{W}\sqrt{\cos2\theta_W}};\\
&&{{g^{Z_He^{+}e^{-}}_L=\frac{eS_W}{2C_{W}\sqrt{\cos2\theta_W}}}},~~~~g^{Z_He^{+}e^{-}}_R=\frac{e(1-3\cos2\theta_W)}{4S_WC_{W}\sqrt{\cos2\theta_W}}; \\
&&V_{Z{\bar{\nu_e}\nu_e}}=\frac{e\gamma_{\mu}P_L}{2C_WS_W},~~~~~~~~~~~~~V_{Z_H{\bar{\nu_e}\nu_e}}=\frac{eS_W\gamma_{\mu}P_L}{2C_W\cos2\theta_W};\\
&&V_{{Z_{{H\mu}}}Z_{{H\nu}}H}=-\frac{e^2fx}{\sqrt{2}C^2_{W}S^2_{W}}g_{\mu\nu},~~{{V_{Z_{\mu}Z_{H_\nu}H}=\frac{e^2fx}{\sqrt{2}C^2_{W}\sqrt{\cos2\theta_W}}}}g_{\mu\nu};\\
&&V_{t\bar{t}\phi^{0}}=-\frac{iy}{\sqrt{2}}S_{L}S_{R},~~~~~~~~~~~~~V_{t\bar{t}H}=-\frac{em_tC_LC_R}{2m_WS_{W}};\\
&&V_{t\bar{T}H}=-\frac{y}{\sqrt{2}}[(C_{L}S_{R}+S_{L}C_{R}x)P_{L}+(C_{L}S_{R}x+S_{L}C_{R})P_{R}];\\
&&V_{\phi^{0}Z_{\mu}H}=\frac{iexp3_{\mu}}{6S_{W}C_{W}},~~~~~~V_{\phi^{0}Z_{H\mu}H}=\frac{iex[(14-17S^2_{W})p2_{\mu}-(4-S^2_{W}p1_{\mu})]}{18S_{W}C_{W}\cos2\theta_W};\\
&&{{V_{Z_{\mu}Z_{H_\nu}HH}=\frac{e^2}{C^2_{W}\cos2\theta_W}}}g_{\mu\nu},~~~V_{{Z_{{H\mu}}}Z_{{H\nu}}HH}=-\frac{e^2}{C^2_{W}S^2_{W}}g_{\mu\nu};
\end{eqnarray}
where
\begin{eqnarray}
S_{L}&=&\frac{1}{\sqrt{2}}\sqrt{1-(y^{2}f^{2}\cos 2x+M^{2})/N_{t}},~~C_{L}=\sqrt{1-S^{2}_{L}};\\
S_{R}&=&\frac{1}{\sqrt{2}}\sqrt{1-(y^{2}f^{2}\cos
2x-M^{2})/N_{t}},~~C_{R}=\sqrt{1-S^{2}_{R}}.
\end{eqnarray}

\section{Higgs productions in the LRTH model at $e^{+}e^{-}$
colliders }\ In this section, we will study the contributions of the
LRTH model to three different types of Higgs boson production
processes at $e^{+}e^{-}$ colliders separately. In our calculations,
the SM input parameters are taken from Ref. \cite{SM-parameter}. We
take the SM-like Higgs mass as $m_H=125$ GeV. The LRTH parameters
involved in the amplitudes are the breaking scale $f$, the masses
$m_{T}$, $m_{Z_{H}}$, $m_{\phi^{0}}$ and the mixing parameter $M$.
The masses $m_{T}$, $m_{Z_{H}}$ and $m_{\phi^{0}}$ are correlated to
$f$ and $M$, and parts of their values are listed in table
\ref{tab:table1}. The value of the mixing parameter $M$ is
constrained by the $Z\rightarrow b\bar{b}$ branching ratio and
oblique parameters \cite{phenomenology-1}. In our analysis, we take
small $M$ and pick two typical values of $M=$ 0 and $M$= 150.
\begin{table}[]
\begin{center}
\caption{ The masses (in GeV) of $m_{T}$, $m_{Z_{H}}$ and $m_{\phi^{0}}$ 
used in this paper. } \label{tab:table1} \vspace{0.1in}
\doublerulesep 0.8pt \tabcolsep 0.1in
\begin{tabular}{|l|l|l|l|l|l|l|l|l|}\hline
       $f$ (GeV) &800&900&1000&1100&1200&1300&1400&1500\\
\hline $m_{T}(M=0)$&783.1&885.5&987.5&1089.2&1190.7&1291.9&1393.0&1494.0\\
\hline $m_{T}(M=150)$&809.8&908.5&1007.4&1106.6&1206.0&1305.4&1404.9&1504.5\\
\hline $m_{Z_{H}}(M=0)$&2307.9&2676.3&3038.5&3396.0&3750.0&4100.9&4449.6&4796.4\\
\hline $m_{Z_{H}}(M=150)$&2403.0&2761.3&3115.1&3465.5&3813.3&4159.1&4503.2&4845.9\\
\hline $m_{\phi^{0}}(M=0)$&113.4&115.2&116.4&117.4&118.1&118.7&119.2&119.5\\
\hline $m_{\phi^{0}}(M=150)$&115.7&117.0&117.9&118.6&119.2&119.6&119.9&120.2\\
\hline
\end{tabular} \end{center} \end{table}

Recently, the ATLAS Collaboration presented the results that a
narrow resonance with SM $Z$ couplings to fermions is excluded at
95\% C.L. for masses less than 2.79 TeV in the dielectron channel,
2.53 TeV in the dimuon channel, and 2.90 TeV in the two channels
combined \cite{ATLS-Z}.  And presented the limit on a
Grand-Unification model based on the $E_6$ gauge group, a spin-2
graviton excitation from Randall-Sundrum models, etc. The same thing
has also been explored by the CMS Collaboration and a sequential SM
$Z'$ resonance lighter than 2.59 TeV \cite{CMS-Z} is excluded at
95\% C.L..

\begin{figure}[htbp]
\begin{center}
\scalebox{0.25}{\epsfig{file=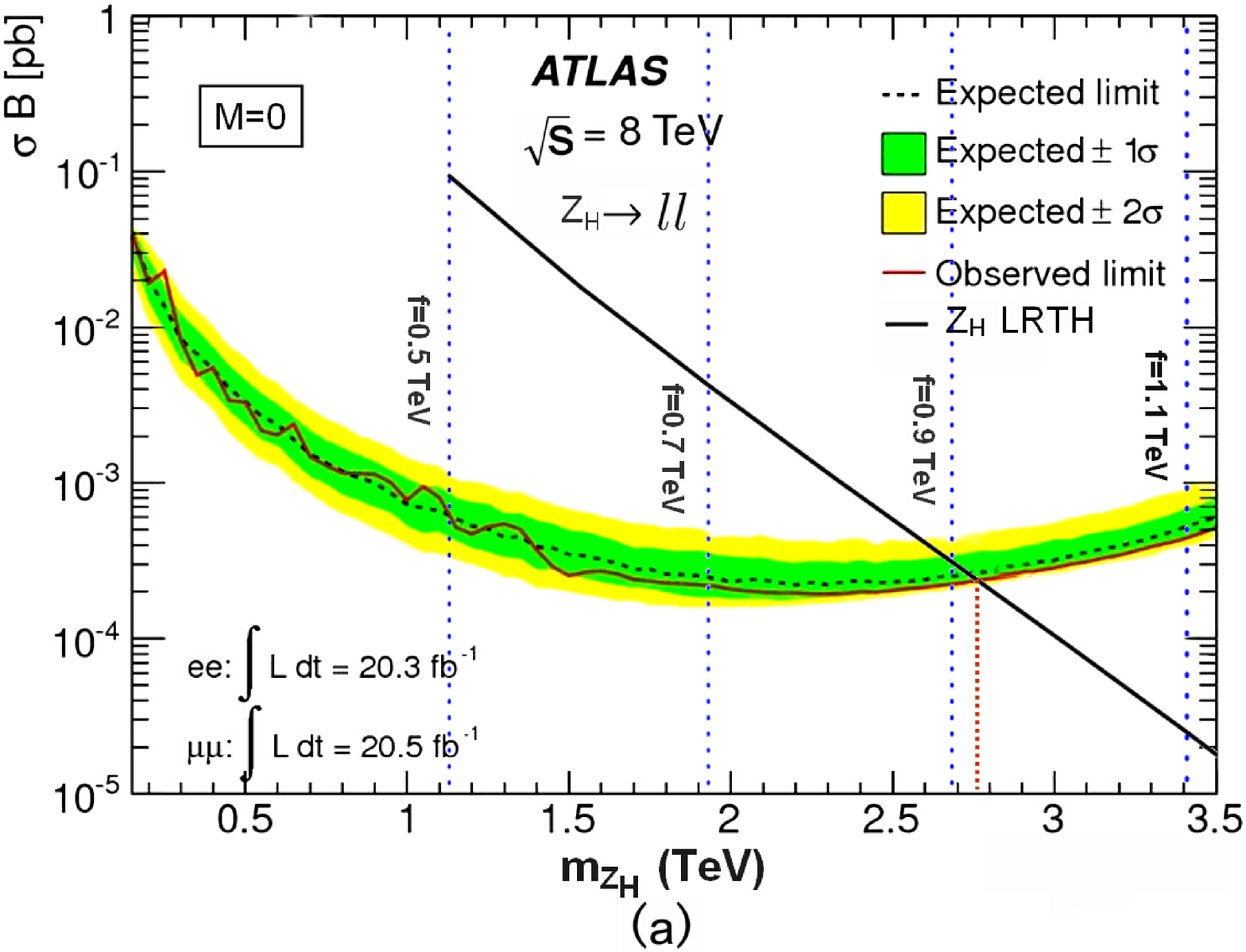}}\hspace{0cm}
\scalebox{0.25}{\epsfig{file=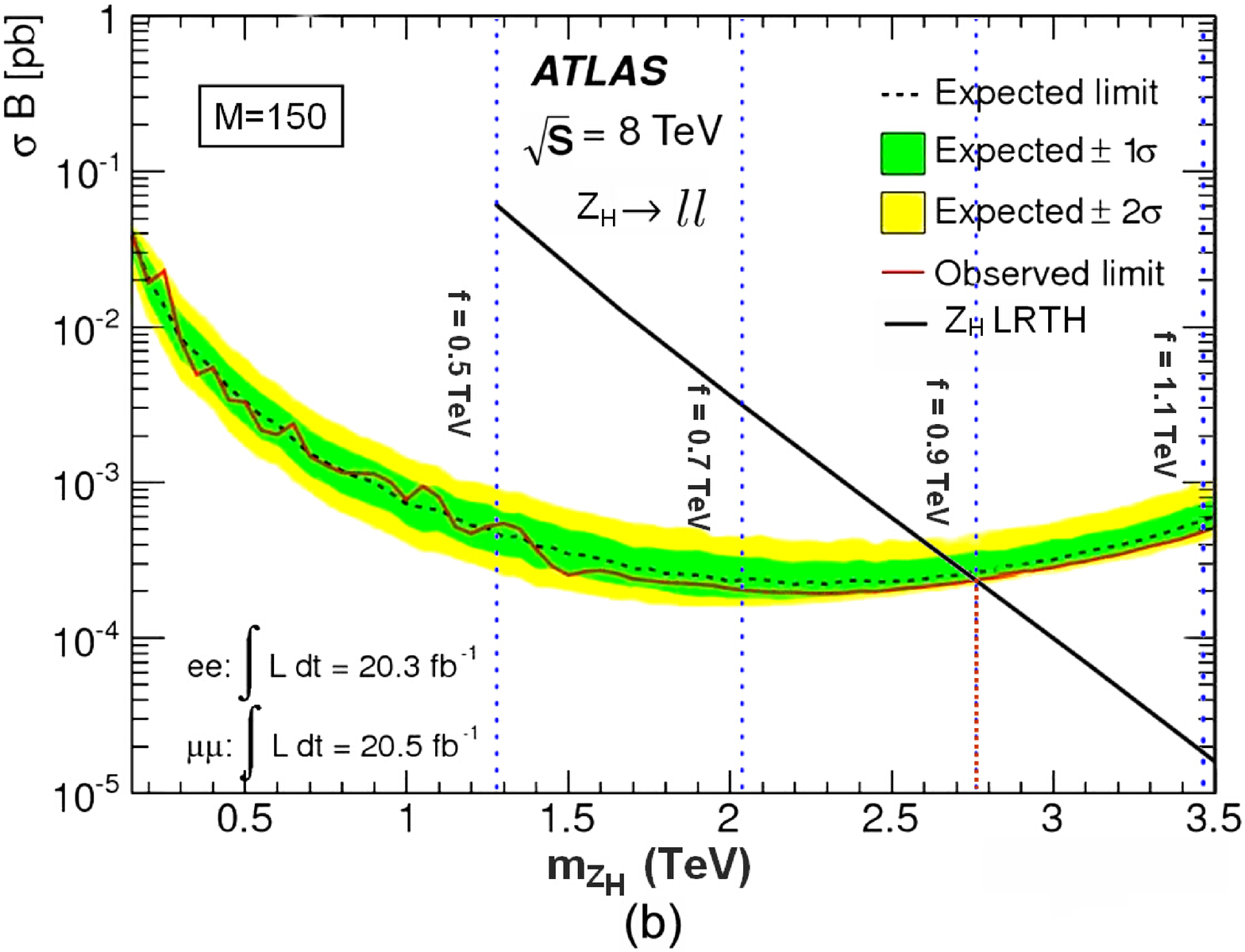}}\vspace{-0.3cm}
\caption{$\sigma (q\bar{q}\rightarrow Z_H)\times Br(Z_H\rightarrow
l^{+}l^{-})$ (where $l=e~{\rm or}~\mu$) as a function of $m_{Z_H}$
at 95\% C.L. observed and expected data at the LHC for M=0 (a) and
M=150 (b) in the LRTH model.}\label{fig:limit}
\end{center}
\end{figure}

In order to constrain the mass of $Z_H$ from the LRTH model, we show
the observed and expected 95\% C.L. exclusion limits on $\sigma
(q\bar{q}\rightarrow Z_H)\times Br(Z_H\rightarrow l^{+}l^{-})$
(where $l=e~{\rm or}~\mu$) as a function of $m_{Z_H}$ at the LHC in
Fig.\ref{fig:limit}, where the observed and expected exclusion
limits come from Ref. \cite{ATLS-Z}. We have checked the production
process $q\bar{q}\rightarrow Z_H$ and the decay $Z_H\rightarrow
l^{+}l^{-}$, and found that our results were consistent with those
in Ref. \cite{phenomenology-1}. From the Fig. \ref{fig:limit}, we
can see that the limits on the $m_{Z_H}$ are insensitive to $M$. In
two cases, the $m_{Z_H}$ are both required to be larger than 2.76
TeV, this is corresponding to the scale $f>$ 920 GeV for $M=0$ and
$f>$ 900 GeV for $M=150$, which are much stronger than the
constraints from the LHC Higgs data \cite{lyb-h-date}.

Meanwhile, there are many searches on the heavy top partners have
been performed by both ATLAS \cite{ATLAS-T-1,ATLAS-T-2} and CMS
\cite{CMS-T-1,CMS-T-2} collaborations. The results show that $T$
quarks with masses below 745 GeV are excluded at 95\% C.L. for
exclusive decays of $T\rightarrow tH$. For different decay modes of
the $T$ quark, the resulting mass limits range from 697 GeV, for
Br($T\rightarrow tZ$) = 20\% and Br($T\rightarrow bW$) = 80\%, to
782 GeV for Br($T\rightarrow tZ$) = 100\%.
However, the top quark parter $T$ in the LRTH model can decay into
$b\phi^{+}$, $bW$, $tH$, $tZ$ and $t\phi^{0}$, and more than 70\% of
heavy top decays via $T\rightarrow b\phi^{+}$. The branching ratios
of the other decays modes are suppressed since the relevant
couplings are suppressed by at least one power of $M/f$. In the
limit $M=0$, the only two body decay mode is $T\rightarrow
b\phi^{+}$ with a branching ratio of 100\% \cite{phenomenology-1}.
Thus, the current constraint on the top partner will be relaxed in
the LRTH model. In addition, we have checked that the limit of the
scale $f>$ 900 GeV satisfies the limit from the searches of $T$
quark.

\subsection{Higgs-Gauge coupling}
\noindent

\begin{figure}[htbp]
\scalebox{0.6}{\epsfig{file=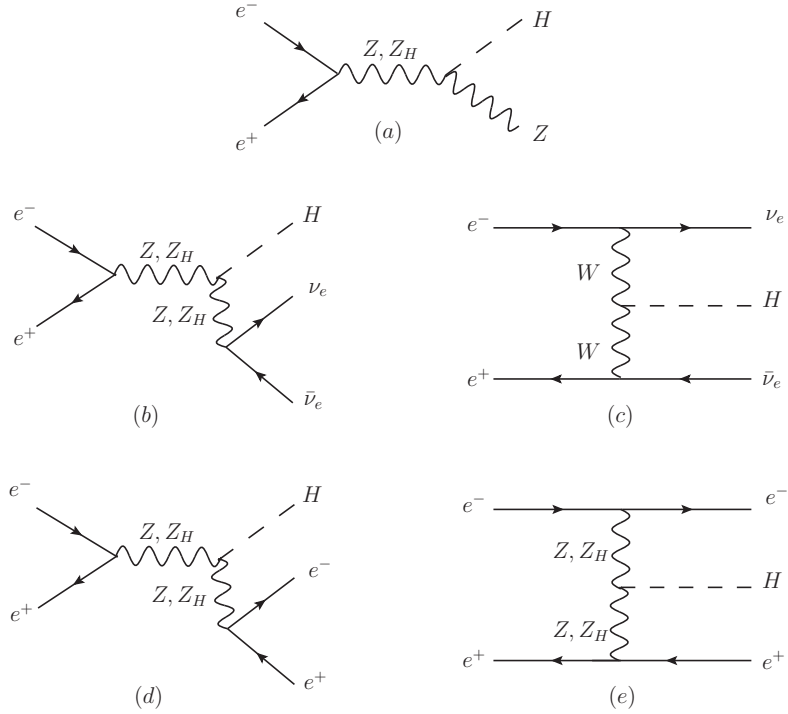}}\vspace{-0.5cm}
\caption{Lowest-order Feynman diagrams for $e^{+}e^{-}\rightarrow
ZH$(a), $e^{+}e^{-}\rightarrow \nu_{e}\bar{\nu_{e}}H$(b,c) and
$e^{+}e^{-}\rightarrow e^{+}e^{-}H$(d,e).}\label{fig:eezh}
\end{figure}

In the LRTH model, the lowest-order Feynman diagrams of the
processes $e^{+}e^{-}\rightarrow ZH$, $e^{+}e^{-}\rightarrow
\nu_{e}\bar{\nu_{e}}H$ and $e^{+}e^{-}\rightarrow e^{+}e^{-}H$ are
shown in Fig. \ref{fig:eezh}. In comparison with the SM, we can see
that the tree-level Feynman diagrams of these processes in the LRTH
model receive the additional contributions arising from the heavy
gauge boson $Z_H$.

\begin{figure}[htbp]
\begin{center}
\scalebox{0.75}{\epsfig{file=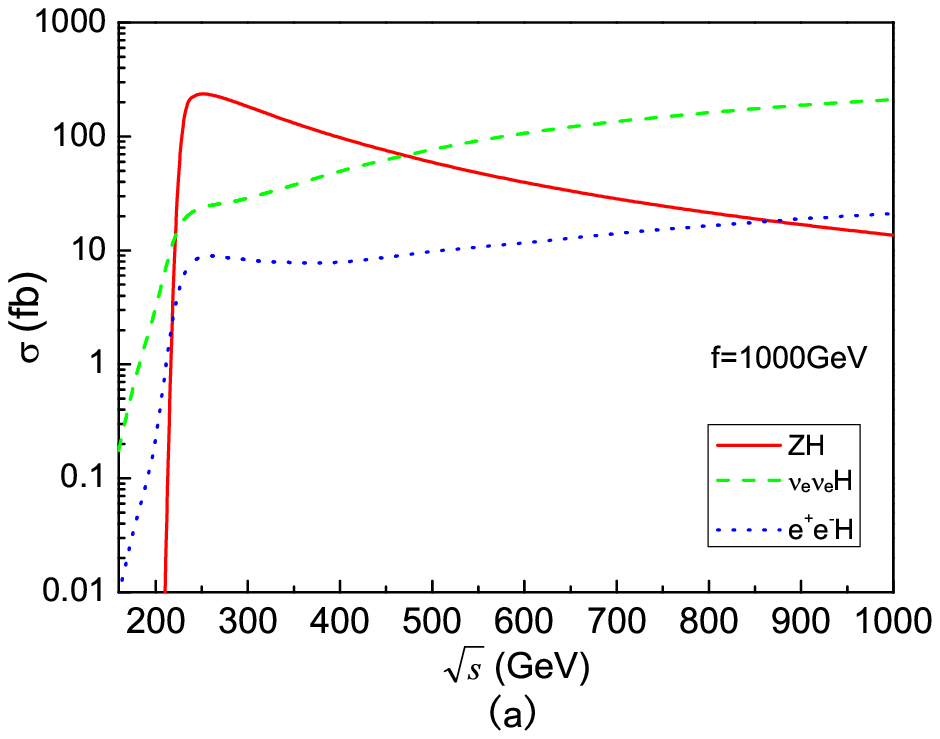}}\hspace{-1cm}
\scalebox{0.75}{\epsfig{file=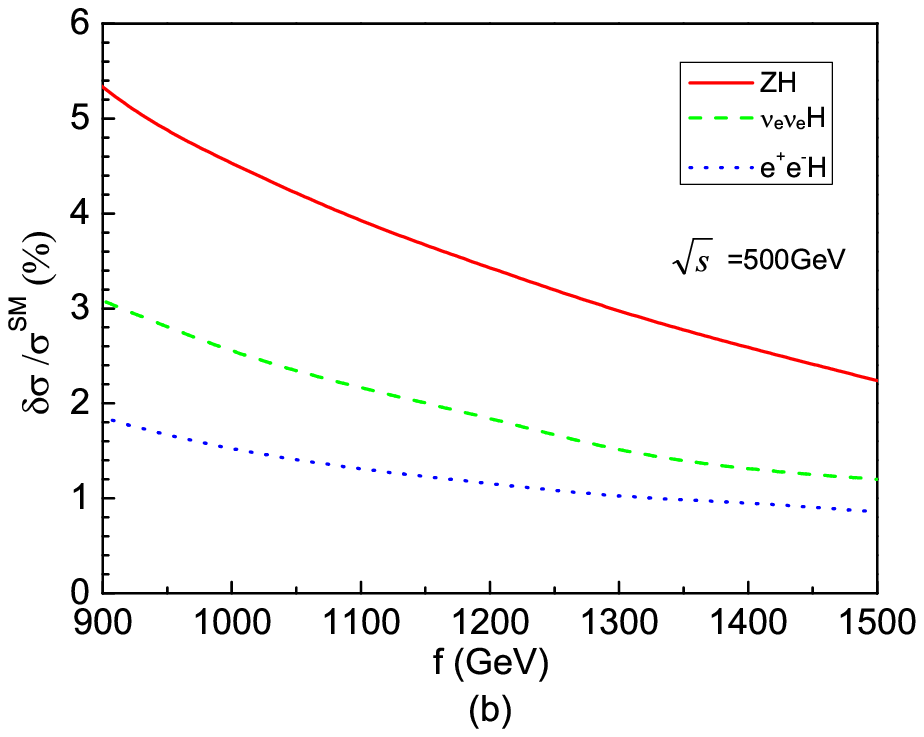}}\vspace{-0.5cm} \caption{The
production cross section $\sigma$ versus the c.m. energy $\sqrt{s}$
for $f=1000$ GeV(a) and the relative correction $\delta
\sigma/\sigma^{SM}$ versus the scale $f$ for $\sqrt{s}=500$ GeV(b)
in the LRTH model.}\label{fig:cross-zh}
\end{center}
\end{figure}

In Fig. \ref{fig:cross-zh}(a), we show the production cross section
$\sigma$ of the three processes as functions of the c.m. energy
$\sqrt{s}$ for the scale $f=1000$ GeV in the LRTH model. We can see
that the Higgs strahlung process $e^{+}e^{-}\rightarrow ZH$ attains
its maximum at $240\sim 250$ GeV, the cross section for $ZH$ process
is in proportion to $1/s$ and dominates the fusion process at the
low energies. While the cross section for $e^{+}e^{-}\rightarrow
\nu_{e}\bar{\nu_{e}}H$ rises as log($s/m^2_H$) and dominates at high
energies. The $\nu_{e}\bar{\nu_{e}}H$ and $e^{+}e^{-}H$ production
cross sections increase with the c.m. energy and can respectively
take over that of the $ZH$ process at $\sqrt{s}\geq 500$ GeV and
$\sqrt{s}\geq 900$ GeV, where the cross section for the process
$e^{+}e^{-}H$ is suppressed by an order of magnitude compared with
the process $\nu_{e}\bar{\nu_{e}}H$.

In Fig. \ref{fig:cross-zh}(b), we show the relative correction
$\delta \sigma/\sigma^{SM}$ of the three production channels as
functions of the scale $f$ for $\sqrt{s}=500$ GeV, respectively,
where $\delta \sigma$ is defined as $\delta
\sigma=\sigma^{LRTH}-\sigma^{SM}$. We can see that the values of the
relative corrections decrease with the scale $f$ increasing, which
indicates that the effects of the LRTH model will decouple at the
high scale $f$. In the same parameter space, the three curves also
demonstrate the process $e^{+}e^{-}\rightarrow ZH$ has the largest
relative correction, which can maximally reach 5.3\% when the scale
$f$ is as low as 900 GeV.

For the process $e^{+}e^{-}\rightarrow ZH$, the 250(500) GeV run of
the ILC  can measure the cross section to a relative accuracy of
2.5(3.0)\% at 250(500) fb$^{-1}$ \cite{ILC-Higgs-White-Paper,
ILC-Report-2}. In addition, an even more remarkable precision of
0.4\% may be achieved at the recently proposed Triple-Large
Electron-Positron Collider (TLEP)\cite{TLEP}, which is a new
circular $e^{+}e^{-}$ collider operated at $\sqrt{s}$=240 GeV with
$10^4$ fb$^{-1}$ integrated luminosity.
 For the process $e^{+}e^{-}\rightarrow
\nu_{e}\bar{\nu_{e}}H$, the ILC can measure this cross section times
the branching fraction to $b\bar{b}$ to a statistical accuracy of
about 0.6\% \cite{ILC-Higgs-White-Paper, ILC-Report-2} at 500 GeV
with an integrated luminosity of 500 fb$^{-1}$.
 For the process $e^{+}e^{-}\rightarrow e^{+}e^{-}H$, we can see that the relative
correction to the cross section of this process is very small in the
LRTH model. Meanwhile, such a process is dominated by the huge SM
backgrounds at the ILC.  So we can conclude that it is not promising
to observe the LRTH effects through $e^{+}e^{-}\rightarrow
e^{+}e^{-}H$, as a comparison with $e^{+}e^{-}\rightarrow ZH$ at the
ILC.
\subsection{Top quark Yukawa coupling}
\begin{figure}[htbp]
\scalebox{0.5}{\epsfig{file=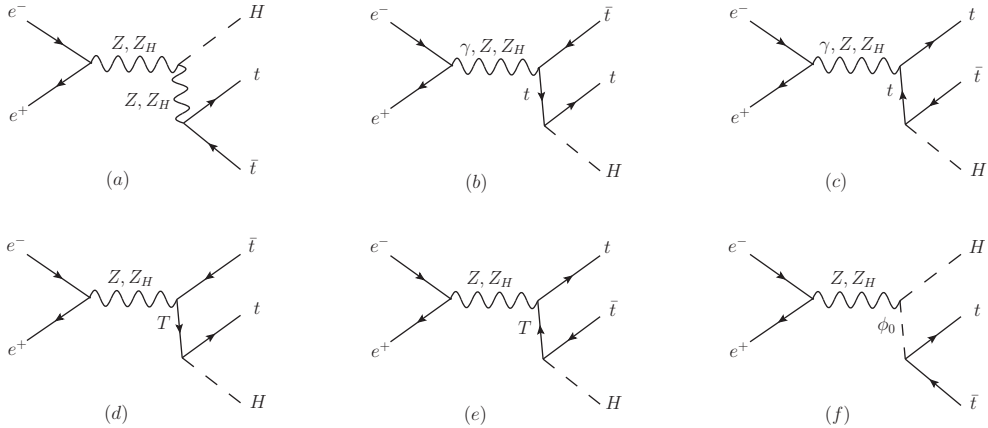}}\vspace{-0.5cm}
\caption{Lowest-order Feynman diagrams for $e^{+}e^{-}\rightarrow
t\bar{t}H$}\label{fig:eetth}
\end{figure}
The relevant tree-level Feynman diagrams of the process
$e^{+}e^{-}\rightarrow t\bar{t}H$ in the LRTH model are shown in
Fig. \ref{fig:eetth}. Comparing with the SM, we can see that there
are additional diagrams mediated by the heavy gauge boson ${Z_H}$,
the heavy $T$-quark and the pseudo-scalar $\phi_0$ in the LRTH
model. Although a contribution can also come from the pseudo-scalar
$\phi_0$, such a contribution is relatively small since the
$\phi_0t\bar{t}$ coupling is suppressed by the factor $(M^4/f^4$).

\begin{figure}[htbp]
\scalebox{0.75}{\epsfig{file=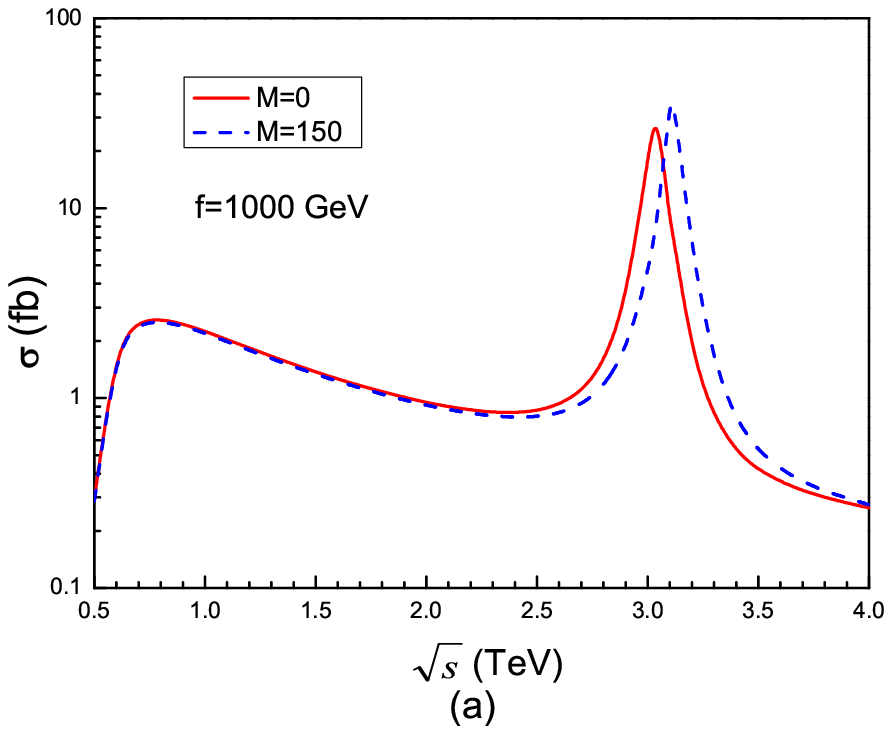}}\hspace{-1cm}
\scalebox{0.75}{\epsfig{file=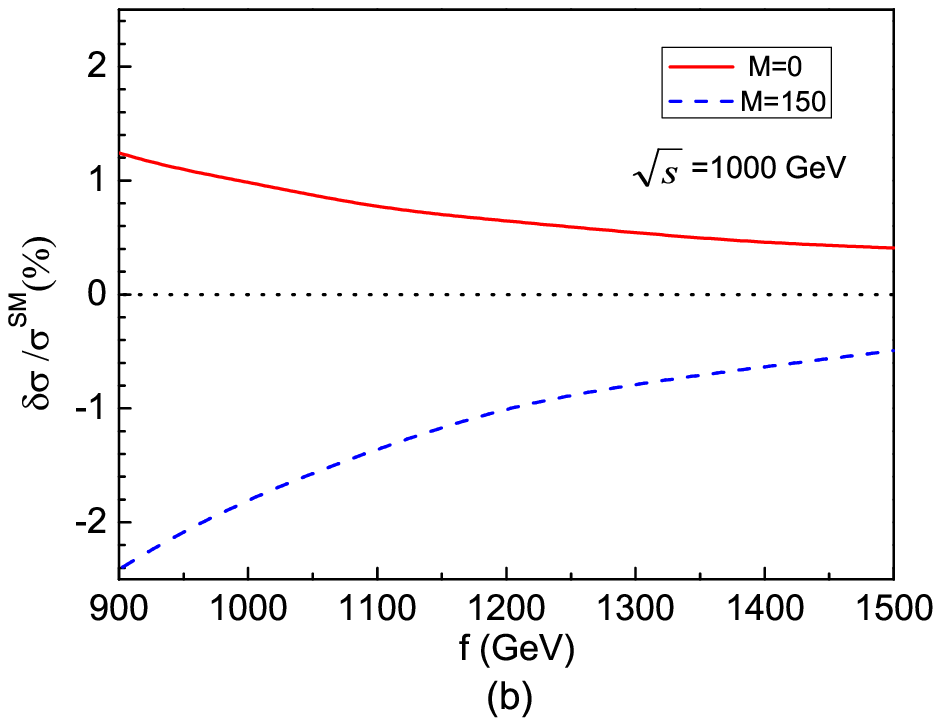}}\vspace{-0.7cm} \caption{The
production cross section $\sigma$ versus the c.m. energy $\sqrt{s}$
(a) and the relative correction $\delta \sigma/\sigma^{SM}$ versus
the scale $f$ for $\sqrt{s}=1000$ GeV (b) in the LRTH
model.}\label{fig:cross-tth}
\end{figure}

In Fig. \ref{fig:cross-tth}(a), we show the production cross section
$\sigma$ as  functions  the c.m. energy $\sqrt{s}$ in the LRTH
model. We take $f=1000$ GeV and $M=0,150$ GeV as examples. Since the
process proceeds mainly through the $s$-channel, we can see that the
cross section resonance emerges when $m_{Z_{H}}$ approaches the
$\sqrt{s}$. For the same scale $f$, the resonance peak for case
$M=0$ is smaller than that for case $M=150$ GeV. However, the
detection for such resonance effect is beyond the reach of the ILC,
this could be accessed later by a multi-TEV $e^{+}e^{-}$ collider
\cite{CLIC}. In Fig. \ref{fig:cross-tth}(b), we show the relative
correction $\delta \sigma/\sigma^{SM}$ of the process
$e^{+}e^{-}\rightarrow t\bar{t}H$ as functions of the scale $f$ for
$M=0,150$ GeV at the ILC. We can see that the deviation is positive
for $M=0$ and the deviation is negative for $M=150$ GeV. When the
scale $f$ ranges from 900 GeV to 1500 GeV, the values of relative
correction is less than 2.4\%.

At the ILC, the 10\% accuracy expected at $\sqrt{s}$ = 500 GeV can
be significantly improved by the data taken at 1000 GeV due to the
larger cross section and the less background from
$e^{+}e^{-}\rightarrow t\bar{t}$. Fast simulations at  $\sqrt{s}$ =
800 GeV showed that we would be able to determine the top Yukawa
coupling to 6\% for $m_H$ = 120 GeV, given an integrated luminosity
of 1000 fb$^{-1}$ and residual background uncertainty of 5\%
\cite{ILC-report-tth-dector}. Full simulations just recently
completed by SiD and ILD showed that the top Yukawa coupling could
indeed be measured to a statistical precision of 4.3\% for $m_H$  =
125 GeV with $\sqrt{s}$ = 1000 GeV and the integrated luminosity of
1000 fb$^{-1}$ \cite{ILC-report-Volume4}. By this token, we can see
that the $t\bar{t}H$ production channel will be hard to be observed
at the ILC.
\subsection{Higgs self-coupling}

\begin{figure}[htbp]
\scalebox{0.5}{\epsfig{file=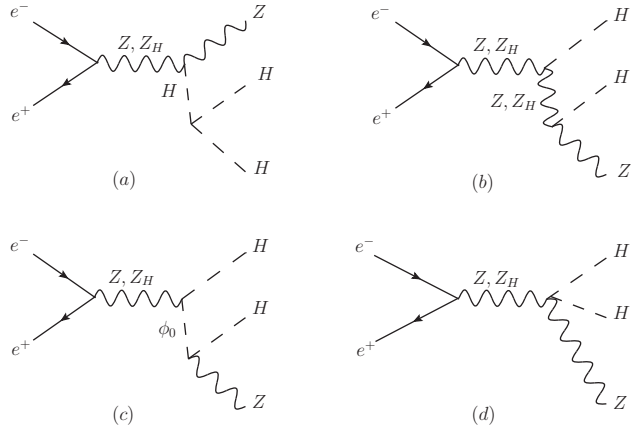}}\vspace{-0.5cm}
\caption{Lowest-order Feynman diagrams for $e^{+}e^{-}\rightarrow
ZHH$}\label{fig:eezhh}
\end{figure}
\begin{figure}[htbp]
\scalebox{0.5}{\epsfig{file=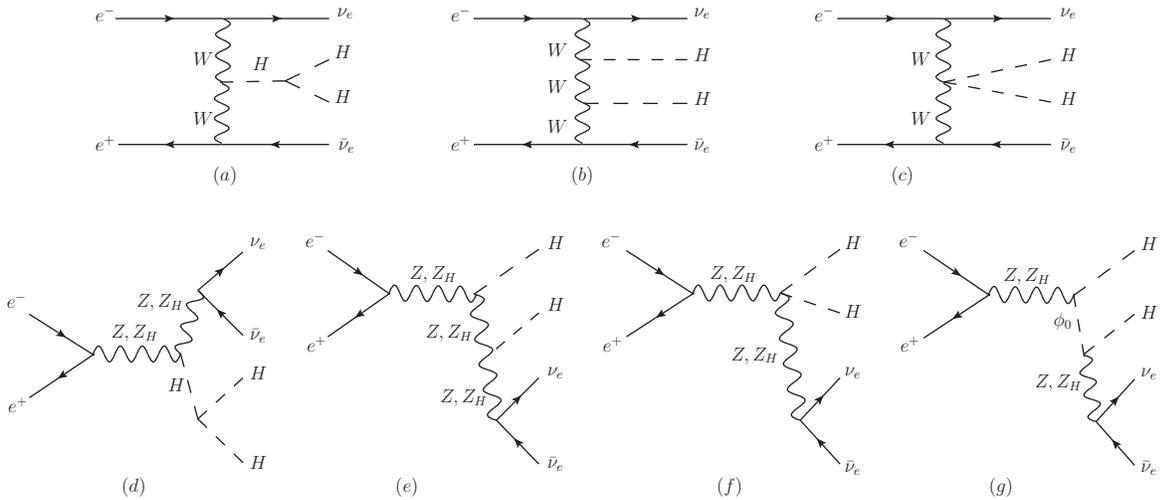}}\vspace{-0.5cm}
\caption{Lowest-order Feynman diagrams for $e^{+}e^{-}\rightarrow
\nu_{e}\bar{\nu_{e}}HH$}\label{fig:eevvhh}
\end{figure}
 \begin{figure}[htbp]
\scalebox{0.95}{\epsfig{file=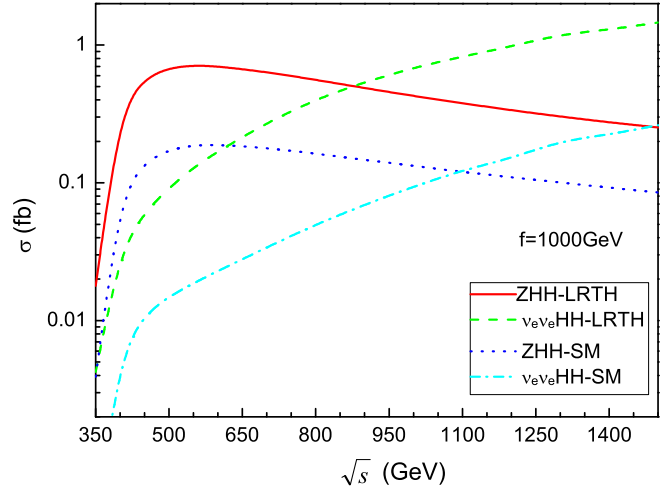}}\vspace{-0.5cm}
\caption{The double Higgs production cross section $\sigma$ versus
the c.m. energy $\sqrt{s}$ for $f=1000$ GeV in the LRTH
model.}\label{fig:cross-zhh}
\end{figure}

In $e^{+}e^{-}$ collisions, the main triple Higgs boson coupling can
be studied through the production channels of double Higgs-strahlung
off $Z$ bosons ($e^{+}e^{-}\rightarrow ZHH$, for $\sqrt{s}$ = 500
GeV) and double Higgs fusion ($e^{+}e^{-}\rightarrow
\nu_{e}\bar{\nu_{e}}HH$, for $\sqrt{s}\geq$ 1 TeV). In the LRTH
model, the relevant Feynman diagrams are shown in Fig.
\ref{fig:eezhh} and Fig. \ref{fig:eevvhh}, respectively. In Fig.
\ref{fig:cross-zhh}, we show the cross sections of the
$e^{+}e^{-}\rightarrow ZHH$ and $e^{+}e^{-}\rightarrow
\nu_{e}\bar{\nu_{e}}HH$ as functions of $\sqrt{s}$ in the SM and the
LRTH model for the scale $f=1000$ GeV. We can see that the cross
section for the former process dominates at the low energies, while
that for the latter process dominates at high energies, and they
have a similar trend in the SM and the LRTH model. Since the Higgs
self-coupling in the LRTH model is quite different from the SM, the
values of cross sections in the LRTH model are much larger than the
SM. The recent studies suggest that a precision of 50\% for the
$HHH$ coupling can be obtained through $pp\rightarrow HH\rightarrow
bb\gamma\gamma$ at the HL-LHC with an integrated luminosity of 3000
fb$^{-1}$ \cite{HL-LHC-1, HL-LHC-2}, and may be further improved to
be around 13\% at the ILC with collision energy up to 1 TeV
\cite{HL-LHC-1}. So, the effects of the LRTH model on these two
processes should be observed at the ILC .
\section{Higgs decay in the LRTH model}

In order to provide more information for probing the LRTH model, we
also give the effect on the Higgs decay.
\begin{figure}[htbp]
\begin{center}
\scalebox{0.7}{\epsfig{file=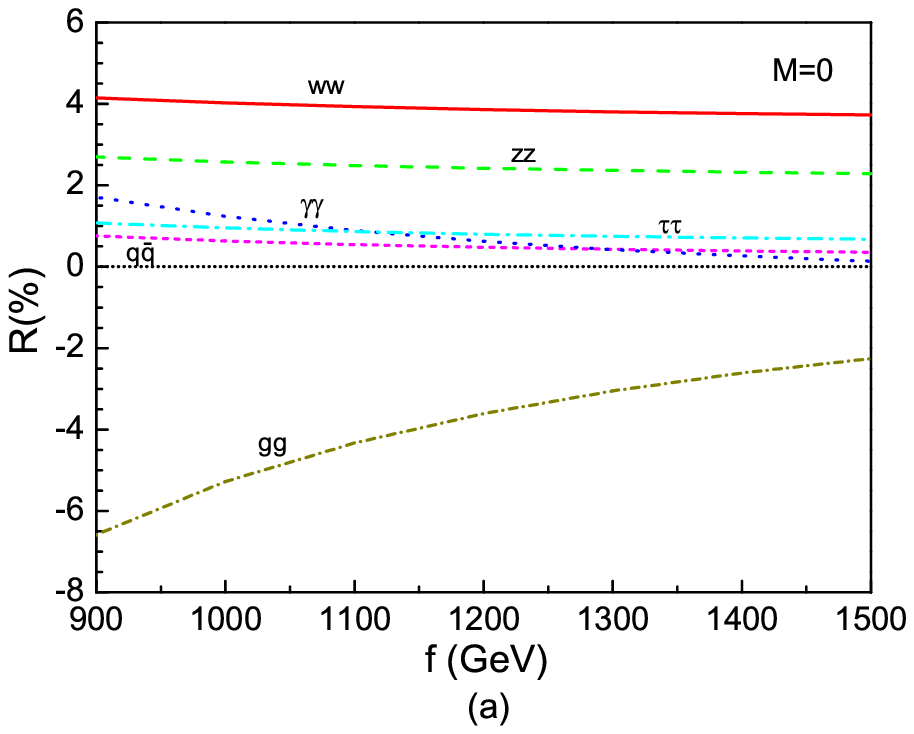}}\hspace{-1cm}
\scalebox{0.7}{\epsfig{file=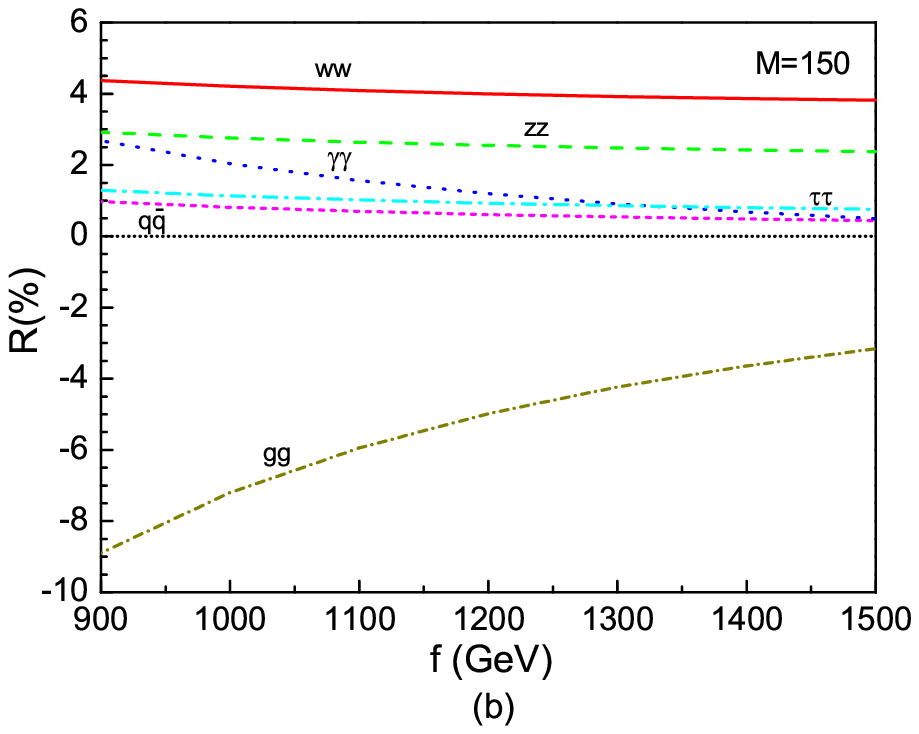}}\vspace{-0.5cm} \caption{ The
relative correction $R$ as a function of the scale $f$ for
$M=0,~150$ GeV in the LRTH model, respectively.}\label{fig:decay}
\end{center}
\end{figure}
In the LRTH model, the major decay modes of the Higgs boson are
$h\rightarrow f\bar{f}$($f=b,c,\tau$), $VV^{*}(V=W, Z)$, $gg$,
$\gamma\gamma$, where $W^{*}/Z^{*}$ denoting the off-shell charged
or neutral electroweak gauge bosons. For $h\to gg$ decay, the LRTH
model can give corrections via the coupling $ht\bar{t}$ and the
heavy top quark loops. For $h\to \gamma\gamma$ decay, the $T-$
quark, $W_H$ boson and $\phi^\pm$ boson loops can provide the
additional contributions simultaneously. By contrast, other decay
models are less affected by the LRTH effect. In our calculations,
the corresponding expressions of decay widths can be found in Refs.
\cite{h-decay-width,lyb-h-date}, the relative correction of the
decay branching ratio is defined by
\begin{equation}
R=(BR^{LRTH}-BR^{SM})/BR^{SM}
\end{equation}

In Fig. \ref{fig:decay}, we show the relative correction $R$ as
functions of the scale $f$ for $M=0,~150$ GeV in the LRTH model. We
can see that the deviation from the SM prediction for $h\to gg$ and
$h\to \gamma\gamma$ decay models decrease and finally reduce to the
SM results with the increasing $f$. The value of relative correction
$R$ for $M=$ 150 is larger than $M=$ 0 and the correction $R_{gg}$
can reach $-8.9\%$. The expected accuracies at the ILC for the
branching ratios of $h\to gg$ are 4.0\% (2.9\%) for $\sqrt{s}$=500
(1000) GeV \cite{ILC-Higgs-White-Paper}, so that the decay mode of
$h\to gg$ might be detected.
\section{Conclusion}
In this paper, we investigated the three types of Higgs bosons
production processes at $e^{+}e^{-}$ colliders under the current LHC
constraints as follows: (i) For the Higgs-Gauge coupling production,
we studied the processes $e^{+}e^{-}\rightarrow ZH$,
$e^{+}e^{-}\rightarrow \nu_{e}\bar{\nu_{e}}H$ and
$e^{+}e^{-}\rightarrow e^{+}e^{-}H$. In the allowed parameter space,
we found that the processes $e^{+}e^{-}\rightarrow ZH$ and
$e^{+}e^{-}\rightarrow \nu_{e}\bar{\nu_{e}}H$ might approach the
observable threshold of the ILC. (ii) For the top quark Yukawa
coupling production process $e^{+}e^{-}\rightarrow t\bar{t}H$, we
found that the deviation of the cross section from the SM prediction
is lower than $2\%$ in a large part of allowed parameter space so
that the effect will be difficult to be observed at the ILC. (iii)
For the Higgs self-coupling production, we studied the processes
$e^{+}e^{-}\rightarrow ZHH$ and $e^{+}e^{-}\rightarrow
\nu_{e}\bar{\nu_{e}}HH$. We found that the cross sections can be
enhanced greatly compared to the SM predictions and these effects
may be observable at the ILC. Besides, we also investigated the
impact of the LRTH model on the Higgs decay and found that the decay
$h\to gg$ had an obvious deviation from the SM prediction.

\section*{Acknowledgement}
We would like to thank Lei Wu for helpful suggestions. This work is
supported by the Joint Funds of the National Natural Science
Foundation of China under grant No. U1404113, by the Science and
Technology Department of Henan province with grant No. 142300410043,
by the National Natural Science Foundation of China under Grant Nos.
11405047, 11305049.

\end{document}